\documentstyle[12pt]{article}
\begin{document}
\title {Remark on the strength of singularities with a $C^0$ metric} 
\author{Amos Ori\\
Department of Physics\\ Technion---Israel Institute of 
Technology\\
32000 Haifa, Israel.}
\date{\today}

\maketitle

\begin{abstract}

Recently Nolan constructed a spherically-symmetric spacetime 
admitting a spacelike curvature singularity with a regular $C^0$ 
metric. We show here that 
this singularity is in fact weak.

\end{abstract}

In a recent paper Nolan \cite{nolan} constructed a 
simple spherically-symmetric spacetime which includes a
spacelike curvature singularity with a continuous ($C^0$) metric. 
The goal was to use this example to 
demonstrate that a curvature singularity with a 
$C^0$ metric may be strong (according to the 
classification by Tipler \cite{tipler} 
and by Ellis and Schmidt \cite{es}).  
In this note we shall show that this singularity is in fact {\it weak}. 
We prove this by solving the second-order differential equation for the 
norm $a$ of the radial Jacobi field, Eq. (n2) (hereafter the letter n 
before the equation number refers to Nolan's paper \cite{nolan}).

We shall use here the notation of Ref. \cite{nolan}. 
It will be assumed that 
the dynamics of $a(t)$ and $x(t)$ is correctly described by the 
corresponding second-order differential equations, i.e. Eq. (n2) for 
$a(t)$ and the equation preceding Eq. (n6) for $x(t)$. 

Since $f=f(x)$ (with $x=u+v$), in Eq. (n2) we substitute 
$f_{uv}=f''$. We first show that
\begin{equation}
a(t)=\dot xe^{-2f}\equiv \bar a(t) \; 
\label{eq1}
\end{equation}
is an exact solution of Eq. (n2). To demonstrate this, we 
differentiate Eq. (\ref{eq1}):
\begin{equation}
\dot a=(\ddot x-2\dot f\dot x)e^{-2f}
=(\ddot x-2f'\dot x^2)e^{-2f} \; ,
\label{eq2}
\end{equation}
where we have used $\dot f=f' \dot x$. From the differential equation 
for $x(t)$ [the one preceding Eq. (n6)] we then find 
\begin{equation}
\dot a=-2f'=-2\dot f/ \dot x \; ,
\label{eq3}
\end{equation}
and hence $\ddot a=-2(\dot f/ \dot x\dot )$ . Now, since
\begin{equation}
f_{uv}=f''=\dot x^{-1}(\dot f/ \dot x\dot ) \; ,
\label{eq4}
\end{equation}
one can easily verify that Eq. (n2) is satisfied.

We now use the Wronskian method to construct a second 
independent solution. Since the Wronskian of Eq. (n2) is a constant, 
this second solution takes the form
\begin{equation}
a(t)=\bar a(t)\int\limits_{}^t {\bar a(t')^{-2}dt'}\equiv \hat a(t) \; .
\label{eq5}
\end{equation}

Consider now the radial Jacobi field which vanishes at $t=t_1$. Its 
norm $a$ is a linear combination of $\bar a(t)$ and $\hat a(t)$ which 
vanishes at $t_1$, so it must take the form
\begin{equation}
a(t)=A\,\bar a(t)\int\limits_{t_1}^t {\bar a(t')^{-2}dt'}\equiv \tilde 
a(t) \; .
\label{eq6}
\end{equation}
Here $A$ is a non-vanishing constant, and without loss of generality we 
may take $A=1$. 

Both $\dot x$ and $e^{-2f}$ are finite and strictly 
positive in the interval $t_1\le t\le 0$,
and so is $\bar a(t)$. 
Consequently, $\tilde 
a(t)$ is finite and non-vanishing everywhere at $t_1< t$, and 
particularly at $t=0$. Thus, the singularity at $t=0$ is weak.

It seems that the error in Ref. \cite{nolan} 
results from a misuse of 
the WKB method in the present case. 
Namely, the inequality before Eq. (n3) does 
{\it not} imply that $a$ will either vanish or diverge at $t=0$. To 
illustrate this by a simple example, consider the equation 
$\ddot a+F(t)\,a=0$ with $a(t)=1+t\,\sin (1/ t)$. Then near $t=0$, 
$a\cong 1$ and
\begin{equation}
F(t)=-\ddot a/ a\cong -\ddot a\cong t^{-3}\sin (1/ t)
\label{eq7} \; ,
\end{equation} 
so the inequality before Eq. (n3) is satisfied 
(to the same extent that it 
is satisfied in Ref. \cite{nolan}; 
i.e. the limit does not exist), and yet $a(t)$ is 
continuous and nonvanishing at $t=0$.

It should be pointed out that Nolan is basically correct in his
claim that Tipler's definition
of weakness is not precisely equivalent to the existence of a
non-singular $C^0$ metric. (The association of weakness with a $C^0$
metric in Ref. \cite{ori92} resulted from a
misinterpretation of a statement in Ref. \cite{tipler}.)
It is not difficult to construct examples of a singular hypersurface
with a non-singular $C^0$-metric, such that the singularity is not
entirely weak.
Such singularities have a more complex structure,
however, and typically
the singularity is strong on subsets of
zero measure only (i.e. on points, lines, or two-surfaces).
The present author is not aware of any example of a
singular hypersurface with a non-singular $C^0$ metric, such
that the singularity is strong in the entire hypersurface
(or even in an open subset of it).

The strength of the null curvature singularity inside a spherical charged 
black hole \cite{ori91}, inside a spinning black hole \cite{ori92}, 
and in the class of solutions constructed 
by Ori and Flanagan \cite{of}, was analyzed independently of the 
continuity of the metric tensor. 
This analysis was based on the 
divergence rate of curvature as a function of {\it proper time}
(as was mentioned explicitly in Ref. \cite{ori91}). 
In all these cases, the singularity was found to be weak
(according to Tipler's definition).
The details of this analysis will be presented in a separate paper.

\end{document}